\documentclass[prl,twocolumn,floatfix,superscriptaddress]{revtex4-1}
\usepackage{dcolumn,amsmath}
\usepackage{graphicx}
\usepackage{bm}
\usepackage{hyperref}

\newcommand{\ecm}{\ensuremath{e {\cdotp} {\rm cm}}}
\newcommand{\eEDM}{{\em e}EDM}
 \newcommand{\Eeff}{\mathcal{E}_\mathrm{eff}}
\newcommand{\B}{\mathcal{B}} %magnetic field
\newcommand{\E}{\mathcal{E}} %electric field
\newcommand{\Bvec}{\vec{\mathcal{B}}} %magnetic field
\newcommand{\Evec}{\vec{\mathcal{E}}} %electric field
\newcommand{\N}{\mathcal{N}} %omega doublet
\newcommand{\Bs}{\tilde{\mathcal{B}}} %magnetic field sign
\newcommand{\Es}{\tilde{\mathcal{E}}} %electric field sign
\newcommand{\Ns}{\tilde{\mathcal{N}}} %omega doublet sign
\newcommand{\muB}{\mu_\mathrm{B}} %Bohr magneton
\newcommand{\vecEeff}{\vec{\mathcal{E}}_\mathrm{eff}}
\newcommand{\de}{d_\mathrm{e}}

\newcommand{\Dp}{D_{\|}}
\newcommand{\Brot}{B_\mathrm{rot}}
\begin{document}
%  \title{Zeeman interaction in ThO $H^3\Delta_1$}
   \title{Zeeman interaction in ThO $H^3\Delta_1$ for the electron EDM search}
\author{A.N.\ Petrov}\email{alexsandernp@gmail.com}
\author{L.V.\ Skripnikov}
\author{A.V.\ Titov}
\affiliation
{Petersburg Nuclear Physics Institute, Gatchina,
             Leningrad district 188300, Russia}
\affiliation{Division of Quantum Mechanics, St.Petersburg State University, 198904, Russia}
\author{N.R.~Hutzler}
\affiliation{Harvard University Physics Department, Cambridge, MA, 02138, USA}
\author{P.W.~Hess}
\affiliation{Harvard University Physics Department, Cambridge, MA, 02138, USA}
\author{B.R.~O'Leary}
\affiliation{Yale University Physics Department, New Haven, CT, 06511, USA}
\author{B.~Spaun}
\affiliation{Harvard University Physics Department, Cambridge, MA, 02138, USA}
\author{D.~DeMille}
\affiliation{Yale University Physics Department, New Haven, CT, 06511, USA}
\author{G.~Gabrielse}
\affiliation{Harvard University Physics Department, Cambridge, MA, 02138, USA}
\author{J.M.~Doyle}
\affiliation{Harvard University Physics Department, Cambridge, MA, 02138, USA}
%\date{}
\begin{abstract}
The current limit on the electron's electric dipole moment, $|\de|<8.7\times 10^{-29}$ \ecm\ (90\% confidence), was set using the molecule thorium monoxide (ThO) in the $J=1$ rotational level of its $H ^3\Delta_1$ electronic state [Science {\bf 343}, 269 (2014)].
This state in ThO is very robust against systematic errors related to magnetic fields or geometric phases, due in part to its $\Omega$-doublet structure.
These systematics can be further suppressed by operating the experiment under conditions where the $g$-factor difference between the $\Omega$-doublets is minimized. 
We consider the $g$-factors of the ThO $H^3\Delta_1$ state both experimentally and theoretically, including dependence on $\Omega$-doublets, rotational level, and external electric field. 
The calculated and measured values are in good agreement. We find that the $g$-factor difference between $\Omega$-doublets is smaller in $J=2$ than in $J=1$, and reaches zero at an experimentally accessible electric field.
This means that the $H,J=2$ state should be even more robust against a number of systematic errors compared to $H,J=1$.
\end{abstract}

\maketitle

%========================================================================
\section{EDM measurements with $\Omega$-doublets}

The experimental measurement of a non-zero electron electric dipole moment (\eEDM, $\de$) would be a clear signature of physics beyond the Standard model
%aptav: Pospelov-Ritz have an important paper in 2014^ it can be included later 
  \cite{Khriplovich1997,Pospelov2005,Feng:2013}.
The most sensitive probes of the electron EDM are precision spin precession measurements in atoms\cite{Regan2002} and molecules\cite{Baron2013,Hudson2011}, which search for energy level shifts resulting from the interaction between the \eEDM\; of a valence electron (or
unpaired electrons) and the large effective internal electric field $\Eeff$\ near a heavy nucleus\cite{Khriplovich1997,Commins2007}.  The current limit, $|\de|<8.7\times 10^{-29}$ \ecm\ (90\% confidence), was set with a buffer-gas cooled molecular beam\cite{Baron2013,Hutzler2012,Patterson2007} of thorium monoxide (ThO) molecules in the metastable electronic $H^3\Delta_1$ state.

Polar molecules have a number of advantages over atoms for electron EDM searches\cite{Commins2010}, including a larger $\Eeff$ and resistance to a number of important systematics.  Some molecules, for example ThO \cite{Vutha2010,Baron2013}, PbO{\cite{DeMille2000,Eckel2013}, HfF$^+$ \cite{Leanhardt2011,Loh2013}, and WC \cite{Lee2009, Lee:13a}, have additional advantages due to the existence of closely-spaced levels of opposite parity, called an $\Omega$-doublet.  Molecules with $\Omega$-doublets can typically be polarized in modest laboratory electric fields ($\lesssim 1-100$ V/cm), and in addition the spin precession measurement can be carried out in a state where the molecular dipole is either aligned or anti-aligned with the external laboratory field. Since $\vecEeff = \Eeff \hat{n}$ points along the internuclear axis, $\hat{n}$,  these states have equal yet opposite projections of $\vec{\E}_\mathrm{eff}$ in the lab frame, and therefore opposite energy shifts due to $\de$.  This means that the experimental signature of $\de$ can be detected either by performing the measurement in the other $\Omega-$doublet state, or by reversing the external electric field $\Evec$.  On the other hand, the internal field of an atom or molecule without $\Omega$-doublets can be reversed only by reversing $\Evec$, which makes the measurement susceptible to systematic errors associated with changing leakage currents, field gradients, and motional fields\cite{Khriplovich1997,Regan2002}.  Molecules with $\Omega$-doublets are very robust against these effects, since the $\Omega$-doublet structure acts as an ``internal co-magnetometer"\cite{DeMille2001}; the spin precession frequencies in the two $\Omega-$doublet states can be subtracted from each other, which heavily suppresses many effects related to magnetic fields\cite{DeMille2001} or geometeric phases\cite{Vutha2009} but doubles the electron EDM signature.  The advantages of $\Omega$-doublets for suppression of systematic effects were first proposed\cite{DeMille2001} and realized\cite{Bickman2009,Eckel2013} by the lead oxide (PbO) electron EDM search.

However, the upper and lower $\Omega$-doublet states have slightly different magnetic $g$-factors, and this difference depends on the lab electric field\cite{Bickman2009}.  Systematic effects related to magnetic field imperfections and geometric phases can still manifest themselves as a false EDM, though they are suppressed by a factor of $\sim\Delta g/g$, where $\Delta g$ is the $g$-factor difference between the two doublet states \cite{Hamilton2010Thesis,Vutha2011Thesis, Eckel2013, Baron2013}. These systematics can be further suppressed by operating the experiment at an electric field where the $g$-factor difference is minimized \cite{Petrov:11, Lee:13a}, or where the $g$-factors themselves are nearly canceled\cite{Shafer-Ray2006}; however, it is clear that understanding the $g$-factor dependence on electric fields is important for understanding possible systematic effects in polar molecule-based electron EDM searches.  Additionally, measurement of $\Delta g$ is a good test of an EDM measurement procedure \cite{Baron2013,Loh2013}.

In this paper we consider the $g$-factors of the ThO $H^3\Delta_1$ state, both theoretically and experimentally, including dependence on $\Omega$-doublets, rotational level, and external electric field.

  \section{Theory}

Following the computational scheme of \cite{Petrov:11}, the $g$-factors of the rotational levels in the $H^3\Delta_1$ electronic state of the $^{232}$Th$^{16}$O molecule are obtained by numerical diagonalization of the molecular Hamiltonian (${\rm \bf \hat{H}}_{\rm mol}$) in external electric $\Evec = \E\hat{z}$ and magnetic $\Bvec = \B\hat{z}$ fields over the basis set of the electronic-rotational wavefunctions $\Psi_{\Omega}\theta^{J}_{M,\Omega}(\alpha,\beta)$.
Here $\Psi_{\Omega}$ is the electronic wavefunction, $\theta^{J}_{M,\Omega}(\alpha,\beta)=\sqrt{(2J+1)/{4\pi}}D^{J}_{M,\Omega}(\alpha,\beta,\gamma=0)$ is the rotational wavefunction, $\alpha,\beta,\gamma$ are Euler angles, and $M$ $(\Omega)$ is the projection of the molecule angular momentum on the lab $\hat{z}$ (internuclear $\hat{n}$) axis.
We define the $g$-factors such that Zeeman shift is equal to
\begin{equation} 
%   E_{\rm Zeeman}(\rm{\E}) = -g(\rm{\E})\mu_B \B M.
   E_{\rm Zeeman} = -g \mu_B \B M.
 \label{Zeem}
\end{equation} 
In other words, we use the convention that a positive $g$-factor means that the projection of the angular momentum and magnetic moment are aligned.
Note that this definition of the $g$-factor for the $J{=}1~H^3\Delta_1$ state differs by a factor
of $-2$ from that given in \cite{Skripnikov:13c}.

In our model the molecular Hamiltonian is written as

\begin{eqnarray}
\nonumber
{\rm \bf \hat{ H}}_{\rm mol}={\rm \bf \hat{H}}_{\rm el}+\Brot { \vec{J}}^2 -2\Brot({ \vec{J}}\cdot{ \vec{J}}^e)  + \\
   \mu_{\rm B}({ \vec{L}}^e-g_{S}{ \vec{S}}^e)\cdot{\Bvec}   -{ \vec{D}} \cdot {\Evec} \ ,
\label{MolHam}
\end{eqnarray}
where ${ \vec{J}}$, ${ \vec{L}}^e$, ${ \vec{S}}^e$, ${ \vec{J}}^e{=}{ \vec{L}}^e{+}{ \vec{S}}^e$
are the electronic$-$rotational, electronic orbital, electronic spin, and total electronic momentum operators, respectively.
${\rm \bf \hat{H}}_{\rm el}$ is the electronic Hamiltonian, $\Brot = 9.76$ GHz\cite{Edvinsson1984} is the rotational constant, $\mu_B$ is the Bohr magneton, and $ g_{S} = -2.0023$ is a free$-$electron
$g$-factor.

Our basis set includes four electronic states. The electronic structure calculations described below show that these states contain the following leading configurations in the $\Lambda\Sigma-$coupling scheme:

%\begin{eqnarray}
\begin{align}
\label{Molbasis}
\nonumber
H^3\Delta_1 & : \left| \sigma \downarrow \delta_2 \downarrow \right|, (T_e=5317~{\rm cm}^{-1})\ ,\\
\nonumber
 Q^3\Delta_2 & : \frac{1}{\sqrt{2}}(\left|\sigma \uparrow \delta_2 \downarrow \right| + \left|\sigma \downarrow \delta_2 \uparrow \right|), (T_e=6128~{\rm cm}^{-1})\ ,\\
 \nonumber
A^3\Pi_{0^+} & : \frac{1}{\sqrt{2}}(\left|\sigma \downarrow \pi_1 \downarrow \right| + \left|\sigma \uparrow \pi_{-1} \uparrow \right|), (T_e=10601~{\rm cm}^{-1})\ ,\\
 %\nonumber
 ^3\Pi_{0^-} & : \frac{1}{\sqrt{2}}(\left|\sigma \downarrow \pi_1 \downarrow \right| - \left|\sigma \uparrow \pi_{-1} \uparrow \right|), (T_e=10233~{\rm cm}^{-1})\ .
%\end{eqnarray}
\end{align}
Here $T_e= \langle\Psi_{\Omega}| {\rm \bf \hat{H}}_{\rm el} |\Psi_{\Omega}\rangle$
are energies of the electronic terms, $\sigma$, $\pi$, and $\delta$ are molecular orbitals; $\sigma$ predominantly consists of the Th $7s$ atomic orbital and $\delta,\pi$ consist predominantly of the Th $6d$ orbital. 
The up (down) arrow means electronic spin aligned (anti-aligned) with the internuclear axis.
$T_e$ is known experimentally for the $H^3\Delta_1, Q^3\Delta_2, A^3\Pi_{0^+}$ states\cite{Huber:79}, but is presently unknown for the $^3\Pi_{0^-}$ state.   
  In our calculation we put $T_e(^3\Pi_{0^-})=10233~{\rm cm}^{-1}$ to reproduce the $\Omega-$doubling\cite{Edvinsson1984},  $a=h\times 186(18)$ kHz, for 
% $H^3\Delta_1$. This is in a reasonable agreement (within the computational accuracy) with our present calculation of the $A^3\Pi_{0^+}\to\ {^3\Pi_{0^-}}$ transition energy,
  $H^3\Delta_1$; this value is within the error bar of our present calculation (described below) of the $A^3\Pi_{0^+}\to\ {^3\Pi_{0^-}}$ transition energy,
$T_e(A^3\Pi_{0^+})-T_e(^3\Pi_{0^-})=569~{\rm cm}^{-1}$.
Provided that the {\it electronic} matrix elements are known, the matrix elements of ${\rm \bf \hat{H}}_{\rm mol}$ between states in the basis set (\ref{Molbasis}) can be calculated with the help of angular momentum algebra \cite{LL77}. The required {\it electronic} matrix elements are

\begin{eqnarray}
 \label{Gpar}
       G_{\parallel} &=&\frac{1}{\Omega} \langle H^3\Delta_1 |\hat{L}^e_{\hat{n}} - g_{S} \hat{S}^e_{\hat{n}} |H^3\Delta_1 \rangle = .0083, \\
 \label{Gperp1}
   G_{\perp}^{(1)} &=& \langle  Q^3\Delta_2  |\hat{L}^e_+ -  g_{S}\hat{S}^e_+|H^3\Delta_1  \rangle = 2.706, \\
 \label{Gperp2}
   G_{\perp}^{(2)} &=& \langle H^3\Delta_1  |\hat{L}^e_+ -  g_{S}\hat{S}^e_+| ^3\Pi_{0^\pm} \rangle = 1.414, \\
 \label{Delt1}
    \Delta^{(1)}  &=&	2\Brot\langle Q^3\Delta_2 |J^e_+ | H^3\Delta_1 \rangle~ =  .882~{\rm cm}^{-1}, \\
 \label{Delt2}
    \Delta^{(2)}  &=&	2\Brot\langle H^3\Delta_1 |J^e_+ | ^3\Pi_{0^\pm}  \rangle~ =  .923~{\rm cm}^{-1}, \\
 \label{dip}
   \Dp &=& \langle H^3\Delta_1 |\hat{D}_{\hat{n}} |H^3\Delta_1 \rangle = 1.67~{\rm a.u.} ,\\
 \label{dip1}
   D_{\perp}^{(1)} &=& \langle  Q^3\Delta_2  |\hat{D}_+  |H^3\Delta_1  \rangle = -0.068~{\rm a.u.}, \\
 \label{dip2}
   D_{\perp}^{(2)} &=& \langle H^3\Delta_1  | \hat{D}_+  | ^3\Pi_{0^\pm} \rangle = 0.693~{\rm a.u.}.
\end{eqnarray}
The molecule-fixed magnetic dipole moment parameter $G_{\parallel}$ is chosen in such a way that 
the mean $g$-factor of the upper and lower states, $\bar{g}(J) = (g_e(J) + g_f(J))/2$,
for $J=1$ exactly corresponds to the experimental datum\cite{Kirilov2013}. The molecule-fixed dipole moment, $\Dp$, is taken from experiment \cite{Vutha:2011}.
The positive value for $\Dp$ means that the unit vector $\hat{n}$ along the molecular axis is directed from O to Th.
Note that $\hat{n}$ is defined backwards with respect to the convention used in \cite{Skripnikov:13c}.
$G_{\perp}^{(2)}$ and $\Delta^{(2)}$
are estimated on the basis of the configurations listed in (\ref{Molbasis}) using only angular momentum algebra. 
The {\sc dirac12} \cite{DIRAC12} and {\sc mrcc} \cite{MRCC2013} codes are employed to calculate the matrix elements (\ref{Gperp1}, \ref{Delt1}, \ref{dip1},\ref{dip2})
and the energy of transition between the $A^3\Pi_{0^+}$ and $^3\Pi_{0^-}$ states.
The inner-core $1s-4f$ electrons of Th are excluded from molecular correlation calculations using the valence (semi-local) version of the generalized relativistic effective core potential method \cite{Mosyagin:10a}. Thus, the outermost 38 electrons of ThO are treated explicitly. 
 For Th we have used the atomic basis set from Ref.~\cite{Skripnikov:13c} (30,20,17,11,4,1)/[30,8,6,4,4,1] in calculations of matrix elements (\ref{Gperp1}, \ref{Delt1}, \ref{dip1}) and the energy of transition between the $A^3\Pi_{0^+}$ and $^3\Pi_{0^-}$ states.   
To calculate the matrix element (\ref{dip2}) the basis set is reduced to (23,20,17,11,3)/[7,6,5,2,1] and 20 electrons are frozen due to convergence problems.  
For oxygen the aug-ccpVQZ basis set \cite{Kendall:92} with two removed g-type basis functions is employed, i.e., we have used the (13,7,4,3)/[6,5,4,3] basis set. The relativistic two-component linear response coupled-clusters method with single and double cluster amplitudes is used to account for electron correlation and transition properties.
To compute the matrix elements of operators $\hat{L}^e_+, \hat{S}^e_+$ in the Gaussian basis set, we have used the code developed in \cite{Skripnikov:13b, Skripnikov:13c, Skripnikov:11a}.

In the framework of second$-$order perturbation theory for the $g$-factors of the $f$ and $e$ states of $ H^3\Delta_1$, $g_f$ and $g_e$ respectively,
as functions of $J$ in the absence of electric field we have \cite{Petrov:11, Lee:13a}:

\begin{eqnarray}
\nonumber
 g_e(J)  = -\frac{G_{\parallel}}{J(J+1)} + \frac{G_{\perp}^{(2)} \Delta^{(2)} }{ T_e(H^3\Delta_1)-T_e(A^3\Pi_{0^+})}  + \\
\frac{G_{\perp}^{(1)} \Delta^{(1)} }{ T_e(H^3\Delta_1)-T_e(Q^3\Delta_2)}  \frac{(J+2)(J-1)}{2J(J+1)}\;,
\label{ge}
\end{eqnarray}
\begin{eqnarray}
\nonumber
 g_f(J)  = -\frac{G_{\parallel}}{J(J+1)} + \frac{G_{\perp}^{(2)} \Delta^{(2)} }{ T_e(H^3\Delta_1)-T_e(^3\Pi_{0^-})}  + \\
\frac{G_{\perp}^{(1)} \Delta^{(1)} }{ T_e(H^3\Delta_1)-T_e(Q^3\Delta_2)}  \frac{(J+2)(J-1)}{2J(J+1)}\;.
\label{gf}
\end{eqnarray}
Because of the small value of $G_\parallel$ in the $H^3\Delta_1$ state, contributions from off-diagonal interactions with the other electronic states included in the 
basis set (\ref{Molbasis}) significantly influence the $g$-factors of $H^3\Delta_1$.
Formally, the interactions with other $\Omega=0^\pm, \Omega=2$ states, not included in this basis, also influence the $g$-factors of $H^3\Delta_1$. Note, however, that if one preserves in the configurations of Eq.~(\ref{Molbasis}) only the leading atomic orbitals of Th,
% when transferring the other contributions to the rest configurations (e.g., applying the one-center basis set expansion on the Th nucleus), the former
they would be the only terms generating
nonzero matrix elements (\ref{Gperp1}-\ref{Delt2}) since the operators treated are radially independent. Therefore, the corresponding matrix elements with $\Omega=0^\pm, \Omega=2$ states not included in the basis set~(\ref{Molbasis}) are several times smaller compared to those in Eqs.~(\ref{Gpar}-\ref{Delt2}), and the matrix elements for higher excited states are suppressed even more.  Since the corresponding contribution to the $g$-factors of $H^3\Delta_1$ appear at higher orders in the perturbation, they are negligible for our treatment.  For highly excited states we have additional suppression due to large energy denominators.  Thus, we expect that inclusion of terms arising only from this truncated basis set should adequately describe the $g$-factors of the $H ^3\Delta_1$ state. 

The external electric field mixes levels of opposite parity (with the same $J$ as well as with $\Delta J = \pm1$) and changes the values of the $g$-factors. 
In the present work we have calculated and measured this effect for the $J=1,2$ states in $ H^3\Delta_1$ for electric fields up to several hundred V/cm. 
The major effects come from mixing the rotational levels of the same electronic states, determined by the body$-$fixed dipole moment (\ref{dip}). 
Since the rotational ($\sim40$ GHz) energy spacing for the $H^3\Delta_1$ state and its distance from other electronic states ($\sim 25$ THz) are much larger than the $\Omega$-doublet 
spacing ( $\sim1$ MHz), there is a range of electric fields where the $e$ and $f$ levels are almost completely mixed ($|d(J)\E|\gg aJ(J+1)$)
while the interactions with other 
rotational and electronic states can be treated as a linear perturbation with respect to $\E$.
For this linear Stark regime the difference between the $g$-factors will be (to a good approximation) a linear function of the external electric field,
with the $g$-factor dependence given by  \cite{Bickman2009}:
\begin{equation}
g(J,\N,\E) = \bar{g}(J) + \eta(J)|\E|\N,
\end{equation}
 where $\N=$sign$(M\Omega\vec{\E}\cdot\hat{z})$.  The quantity $\N$ refers to the molecular dipole either being aligned ($\N=+1$, lower energy) or anti-aligned 
($\N=-1$, higher energy) with $\vec{\E}$, $\bar{g}(J)$ is the mean $g$-factor of the upper and lower states, and $\eta$ is a constant which depends on the molecular electronic and rotational state.
Note, that $g_e(J) = g(J,\N=-1,|\E| \rightarrow 0)$ and $g_f(J) = g(J,\N=+1,|\E| \rightarrow 0)$.
Below, for brevity, we will use this relation for non zero lab electric field as well.

%------------------------------------------
%\section{Electronic structure calculations}

\section{Measurement of $g$ and $\eta$}

We write the energy shifts for the $M=\pm 1$ Zeeman levels in the $H$ state in the linear Stark regime as 

\begin{multline}
E  =  -M g(J,\E,\N)\muB\B -\frac{\Dp M\Omega}{J(J+1)}\E  - M\Ns\Es\Eeff\de \\
 =  -M \bar{g}(J)\muB\B -\eta(J)\Ns M \muB|\E|\B - \Ns d(J)|\E|  \\
- M\Ns\Es\Eeff\de. 
\label{hamiltonian}
\end{multline}
From left to right, these terms represent the Zeeman shift, electric field dependence of the magnetic $g$-factors, the DC Stark shift, 
and the electron EDM interacting with the effective internal electric field.  Here $\de$ is the electron EDM, 
$\Eeff=84$ GV/cm\cite{Skripnikov:13c} is the internal effective electric field, and $\muB$ is the Bohr magneton.  
A tilde over a quantity indicates the sign $(\pm 1)$ of a quantity which is reversed in the experiment, 
$\tilde{\B}=$sign$(\vec{B}\cdot\hat{z}),\tilde{\E}=$sign$(\vec{\E}\cdot\hat{z})$ and $\Ns=\N$ for consistency.

As discussed in detail elsewhere\cite{Vutha2010,Campbell2013,Baron2013}, the terms in eq. (\ref{hamiltonian}) are determined by performing a spin-precession measurement on a pulsed molecular beam of ThO molecules.  By measuring the phase accumulated by a superposition of the $M=\pm 1$ Zeeman sublevels (in any level with $J\geq 1$),  we can determine the the spin precession frequency $\omega = \Delta E/\hbar$, where $\Delta E$ is the energy splitting between the $M=\pm 1$ states, and then calculate $\Delta E$.  By measuring this frequency with all possible values of $\Ns,\Es$, and $\Bs$, we can determine each of the terms in eq. (\ref{hamiltonian}) individually.  Specifically, we measure the component of $\omega$ which is either even or odd under reversal (or ``switch") of $\Ns,\Es,$ and $\Bs$.  We denote these components with a superscript indicating under which experimental switches the component is odd; for example, $\omega^{\N\B}$ is the component of the spin precession frequency which is odd under reversal of $\N$ and $\B$, but not $\E$.  For the terms in eq. (\ref{hamiltonian}), we have

\begin{eqnarray}
\hbar\omega^\B & = & -\bar{g}(J)\muB|\B| \\
\hbar\omega^{\N\B} & = & -\eta(J)\muB|\E\B| \\
\hbar\omega^{\N\E} & = & -\de\Eeff.
\end{eqnarray}

\noindent The Stark interaction is a common-mode shift which does not cause spin precession.  All measurements are performed in the $M=\pm 1$ states of either $J=1,2,3$ in $H$, since our measurement scheme relies on driving a $\Lambda$-type transition to an $M=0$ level in the excited electronic $C$ state.  Population is transferred to the $H,J=1,2,3$ states by optically pumping through the $A^3\Pi_{0+}$ electronic state.  To populate $H,J=1$ we pump through the $A,J=0$ state, which can only decay to the $J=1$ state in $H$ since there is no $H,J=0$ state.  To populate the higher rotational levels we pump into higher rotational states in $A$, which reduces our population transfer efficiency and signal sizes; this limited the number of rotational levels which we were able to probe.

\subsection{Measurement of $\eta$}

\begin{table}
\centering
\caption{Measured values of $\eta(J)$ (in units of nm/V) in different electric and magnetic fields.  We expect $\eta(J)$ to be independent of $\E$ and $\B$.  Error bars are a quadrature sum of the 1 $\sigma$ Gaussian statistical uncertainty and the systematic uncertainty discussed in the text. $\eta(3)$ was not measured due to small signal sizes.}
\begin{tabular}{cccc}

$\E$ [V/cm] 	& $\B$ [mG] & $\eta(1)$ & $\eta(2)$ \\ \hline
   36 &   19 &  $-0.81(2)$ 	& -- \\

   36 &  	38 &  $-0.79(2)$	& -- \\

  141 &   19 &  $-0.80(1)$ 	& -- \\

  141 &   38 &  $-0.80(1)$ 	& -- \\

  141 &   59 &  $-0.78(2)$ 	& -- \\	

	106 & 	38 & -- 					& $+0.03(2)$ \\
\hline
\multicolumn{2}{c}{Weighted mean}  &  $-0.79(1)$  &  $+0.03(2)$
\end{tabular}
\label{table-eta}
\end{table}

We can extract $\omega^{\N\B}$ from our data (using the same methods by which we extract $\omega^{\N\E}$ to determine $\de$\cite{Baron2013}), and use the known $\E$ and $\B$ fields to determine the value of $\eta$, via

\begin{equation}
\eta = -\frac{\hbar\omega^{\N\B}}{\muB|\E\B|}. \label{eta-meas}
\end{equation}

With the exception of the $\B=59$ mG and $J=2$ measurements in Table \ref{table-eta}, we determined $\eta$ from the same data set which was 
used to extract $\de$.  By measuring $\eta$ for several values of $|\E|$ and $|\B|$, we ensure that the 
value of $\eta$ is indeed a constant, independent of the applied fields.

\begin{figure}[htbp]
\centering
\includegraphics[width=0.5\textwidth]{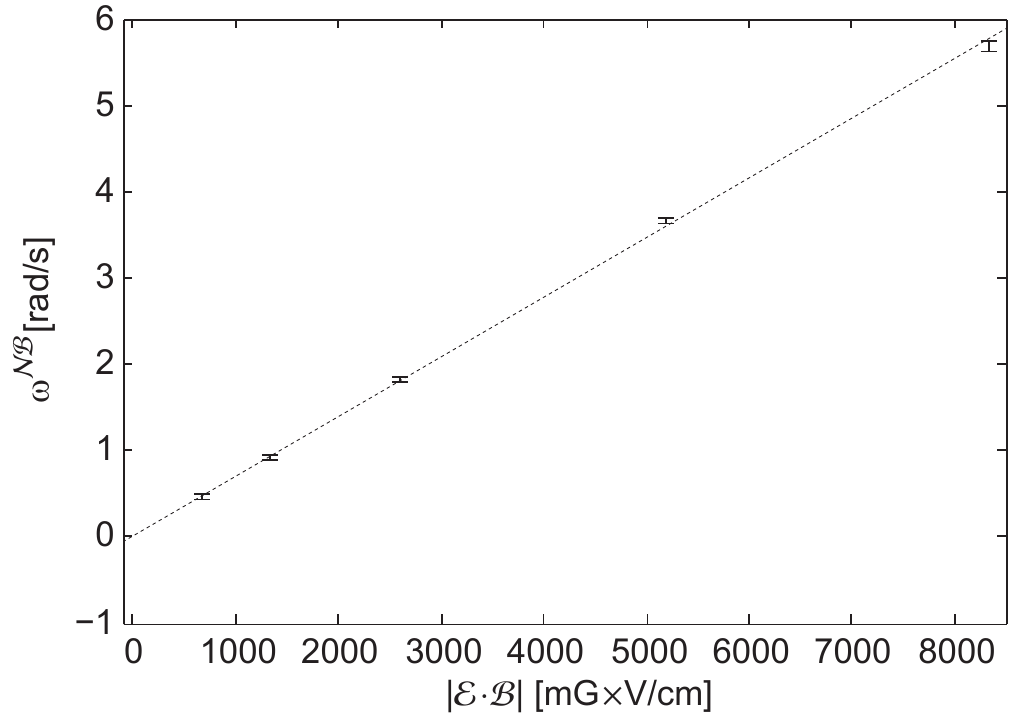}
\caption{Plot of $\omega^{\N\B}$ vs. $|\E\cdot\B|$ for $J=1$ with a linear fit.  According to eq. (\ref{eta-meas}) this slope is $\omega^{\N\B}/|\E\B|=-\eta(1)\muB/\hbar$, from which we extract $\eta(1) = -0.79(1)$.  Error bars are combined statistical and systematic, as in Table \ref{table-eta}.  The reduced $\chi^2$ value of the fit is 1.5, which agrees with the expected value of $1\pm 0.7$ for 4 degrees of freedom.}
\label{fig:omega_nb_vs_eb}
\end{figure}

The uncertainty on $\eta$ comes from a combination of statistical uncertainty on $\omega^{\N\B}$, and from a systematic uncertainty.  The primary systematic error is similar to one affecting our electron EDM measurement, which is discussed in more detail in Ref. \cite{Baron2013}.  Specifically, here an $\N$-correlated laser detuning $\delta^\N$ (caused by differences between the Stark splitting and acousto-optical modulator frequencies used to shift the lasers into resonance) and overall detuning $\delta^{(0)}$ couple to an AC Stark shift to cause a spin precession frequency $\hbar\omega^{\N\B} \propto \delta^{(0)}\delta^{\N}|\B|$.  Since we determine $\eta$ from $\omega^{\N\B}$, this will systematically change our determination of $\eta$.  In the course of the systematic error analysis of our EDM search\cite{Baron2013}, we experimentally measured that $\eta^\mathrm{meas}/(\delta^{(0)}\delta^{\N}) = 2.61(2)$ nm V$^{-1}$ MHz$^{-2}$ with $|\E| = 141$ V/cm, where $\eta^\mathrm{meas}$ is the value of $\eta$ calculated from Eq. (\ref{eta-meas}) by ignoring the AC Stark shift.  Given our measured average $\delta^{(0)}_\mathrm{RMS}\approx 70$ kHz and $\delta^{\N}_\mathrm{RMS}\approx 20$ kHz, this gives rise to a systematic uncertainty in $\eta$ of $\approx 0.01$ nm/V, which is comparable to the statistical uncertainty.  The values of $\E$ and $\B$ are known to $\sim 10^{-3}$ fractionally\cite{Baron2013}, so we do not include those uncertainties in our error budget.

\subsection{Measurement of the $g$-factors}

The measurement of $\bar{g}(1)$ was performed in a previous publication\cite{Kirilov2013}, and we use the value reported there of $\bar{g}(1) = -0.00440(5)$.  The previous measurement did not determine the sign, but the spin precession measurement employed here is sensitive to signs and we find $\bar{g}(1)<0$ (that is, the magnetic moment and angular momentum are anti-aligned in the molecule).

To measure the $g$-factor in the higher rotational $(J)$ levels, we find the smallest magnetic field which results in a $\pi/4$ phase rotation of each Zeeman sublevel.  Because our spin precession measurement is time-resolved, we choose the magnetic field $\B_J$ which results in a $\pi/4$ rotation for the molecules in the center of the beam pulse.  We measure that $\B_J$ =  $19.7, 29.6, 35.5$ mG for $J=1,2,3$ is required to impart a $\pi/4$ phase.

In terms of the flight time $\tau$, the fields $\B_J$ are given by $\bar{g}(J)\muB\B_{J}\tau = \pi/4$.  If we make the assumption that $\tau$ ($\approx 1.1$ ms) does not change during the time it takes to change the lasers to address/populate the other rotational levels, we can see that $\bar{g}(J)/\bar{g}(J') = \B_{J'}/\B_J$ for any $J,J'$.  Since $\bar{g}(1)$ is known, we can solve for $\bar{g}(J) = \bar{g}(1)\times(\B_J/\B_1)$ with the values reported above.  To compute an uncertainty, we make use of the fact that $\tau$ is typically observed to drift on the $\pm 1\%$ level for short time scales, and that the magnetic fields were only set with a resolution of $0.7$ mG.  Together, this gives an overall uncertainty on the $g$-factor measurements (for $J>1$) of $\approx \pm 3\%$.

\section{Results and discussion}
 
Table \ref{gfge} lists the measured and calculated (using Eqs.~(\ref{ge},\ref{gf})) $g$-factors for the $ H^3\Delta_1$ 
for different quantum numbers $J$. 
For a pure Hund's case (a) molecule, we expect $\bar{g}(J) = -G_\|[J(J+1)]^{-1}$\cite{Herzberg1989}.  
However, from comparison of the experimental results (final column) to this expectation (first column) shown in Table \ref{gfge}, we see that this scaling is badly violated. 
Accounting for the contribution of interaction with $Q^3\Delta_2$ (second column of Table \ref{gfge}) leads to much better agreement between the measured and calculated values. 
Furthermore, accounting for perturbation from the $^3\Pi_{0^\pm}$ states makes the agreement better still (third through fifth columns).
 $Q^3\Delta_2$ is the nearest state to $H^3\Delta_1$ and its contribution
 is about an order of magnitude larger compared to  those from the $^3\Pi_{0^\pm}$ states. 
 Note, however, that the interaction with $^3\Delta_2$ (as opposed to the interaction with $^3\Pi_{0^\pm}$) 
 does not contribute in the leading order (at zero electric field) to the difference in $g$-factors of the $f$ and $e$ states.

\begin{table}
\caption{ 
The $g$-factors (in units 10$^{-3}$) calculated and measured for the $H^3\Delta_1$ state in $^{232}$Th$^{16}$O.}
\begin{tabular}{ccccccc}
    \multicolumn{6}{c} {Calculation, Eqs.(\ref{ge},\ref{gf})} & Exper. \\
  J     & $\bar{g}$ 
\footnote[1]{Results when interactions with both $^3\Pi_{0^\pm}$ and $^3\Delta_2$ were omitted. In this case the $g$-factors for $e$ and $f$ states are equal and given by  $-G_{\parallel}/J(J+1)$.}
&    $\bar{g}$ 
\footnote[2]{Results when interactions with only $^3\Pi_{0^\pm}$ were omitted. In this case the $g$-factors
for $e$ and $f$ states are equal.}
 &   $g_f$              &     $g_e$  &  $\bar{g}$ 
\footnote[3]{Results when the parameter $G_{\parallel}$ was chosen in such a way that $\bar{g}(1)$ exactly corresponds to
experimental value. } & $\bar{g}$  \\
\hline
1 &  -4.144  & -4.144 &  -4.409  & -4.391  & -4.400  & -4.40(5)\cite{Kirilov2013} \\
2 & -1.381   & -2.362 &  -2.628  & -2.609  & -2.618  & -2.7(1)\phantom{0[29]} \\
3 &  -0.691  & -1.917 &  -2.182  & -2.164  & -2.173  & -2.4(2)\phantom{0[29]}  \\   
\hline
\end{tabular}
\label{gfge}
\end{table}

In Fig.~\ref{gfgecross} the calculated $g$-factors for the $J = 1,2$ levels of ThO $H^3\Delta_1$ state are shown as functions of the laboratory electric field. 
  Since the electric field mixes $e$ and $f$ levels one might expect that the initial small difference between $g_e$ and $g_f$ would converge to zero with increasing the electric field.
Fig.~\ref{gfgecross}, however, shows that $g_e$ and $g_f$ for $J=1$ do not tend to coincide.
This fact is explained by perturbations from the $J=2$ level, as discussed in \cite{Bickman2009,Petrov:11,Lee:13a}. In turn, the nearest perturbing state for $J=2$ is $J=1$. The energy denominator for $J = 2$ level in the perturbation theory will have the opposite sign compared to the $J = 1$ level and the corresponding curves for $g_e$ and $g_f$ cross each other. 

In Fig.~\ref{eta12} the calculated and experimental values for $\eta(1)$ and $\eta(2)$ are shown. 
For small electric fields, $\eta(J)$ is a function of the electric field which converges to a constant value as the electric field increases.
Both theoretical and experimental data show that for $\E > 36$ V/cm $\eta(1)$ can be considered as independent of $\E$ within experimental
accuracy.

In their search for the electron EDM in the PbO molecule, Bickman \emph{et al}.\cite{Bickman2009} observed dependence of the molecular $g$-factor on the lab electric field $\E$, and found that $\eta(1) = \bar{g}(1)\Dp/(20\Brot)$.
In the ThO $H^3\Delta_1,v=0,J=1$ state, we have $\bar{g}(1) = -0.00440(5)$\cite{Kirilov2013}, $\Dp = h\times 2.13$ MHz/(V/cm)\cite{Vutha2010} and $\Brot = 9.76$ GHz\cite{Edvinsson1984}, and would therefore expect $\eta(1)\approx -1.4$ nm/V based on the treatment from Ref. \cite{Bickman2009}.  Instead we measure $\eta(1)=-0.79(1)$ nm/V, as shown in Table \ref{table-eta}.
The discrepancy is due to the fact that $\bar{g}(J)$ and $\eta(J)$ are much smaller in ThO than in PbO, and therefore the small perturbations from nearby electronic states considered in this paper are of comparable size to the residual values from the mechanisms considered in \cite{Bickman2009}.  

If the magnetic interaction with $^3\Pi_{0^\pm}$ is neglected, then  $g_e = g_f$ for zero electric field and 
mixing between $e$ and $f$ (with the same $J$) does not influence the $g$-factors. 
In this case $\eta(J)$ is a linear function for both small and large electric fields (see dotted (red) curves in Fig.~\ref{eta12}). Similar to case for the zero-field $g$-factor values, the Zeeman interaction with other electronic states has a large contribution to $\eta(J)$, and including this effect makes the measured and predicted values of $\eta(J)$ much closer. 
Due to a large energy separation between different electronic states, the Stark interaction between electronic states (\ref{dip1},\ref{dip2}) 
has smaller effects on the $g$-factors of $H^3\Delta_1$. We have found, however, that it is not negligible; taking this interaction into account significantly improves the agreement between experimental and theoretical values, particularly for $\eta(J=2)$.

The small value of $\eta(2)$ means that the $H,J=2$ state should be even more robust against a number of systematic errors, as compared to $H,J=1$.  
Since the energy shift due to $\de$ does not depend on $J$ when the molecule is fully polarized\cite{Kozlov2002}, performing an EDM measurement 
in multiple rotational levels could be a powerful method to search for and reject systematics in this type of $\Omega$-doublet system.

\begin{figure}[pH]
\includegraphics[width = 3.3 in]{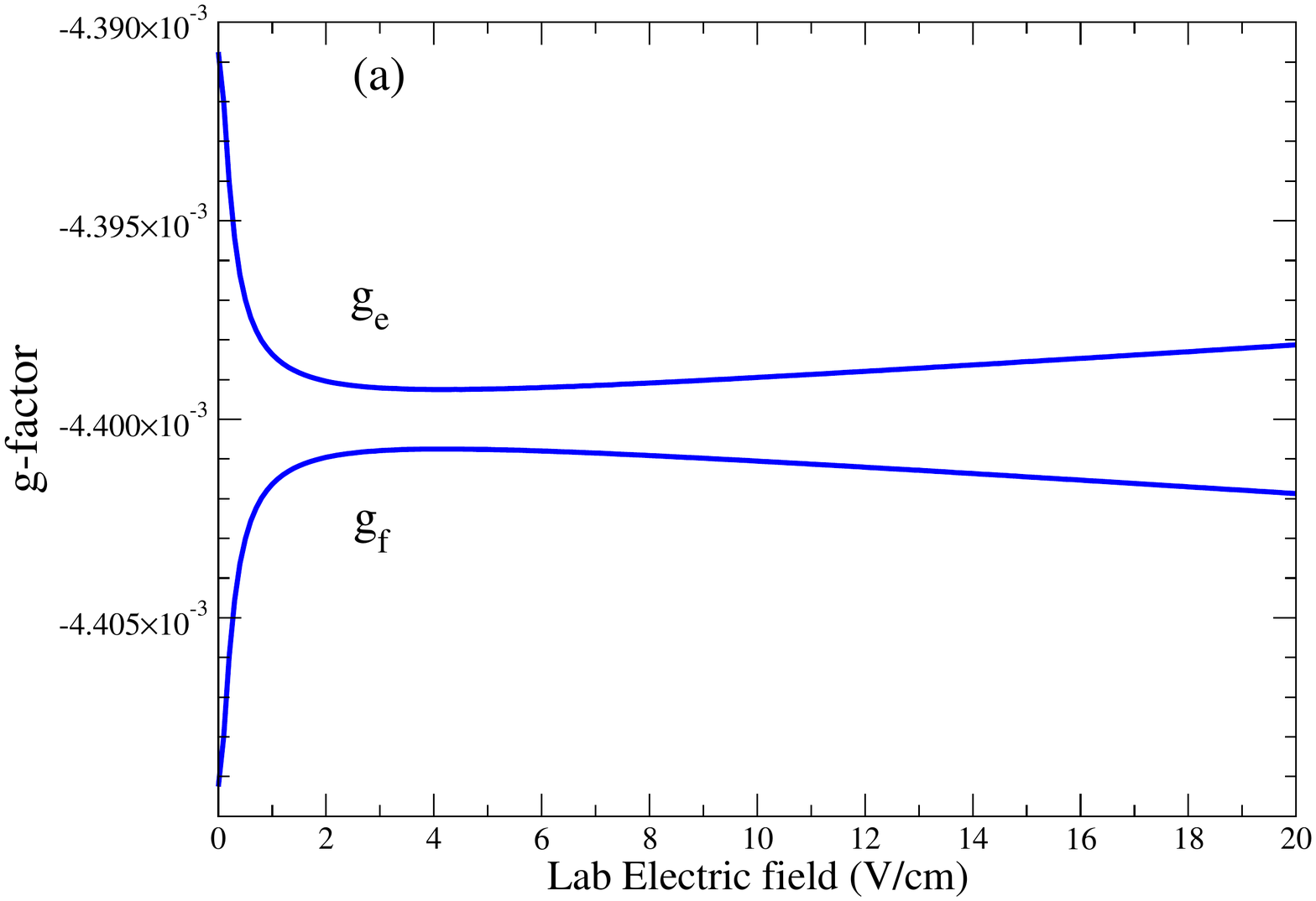}
\includegraphics[width = 3.3 in]{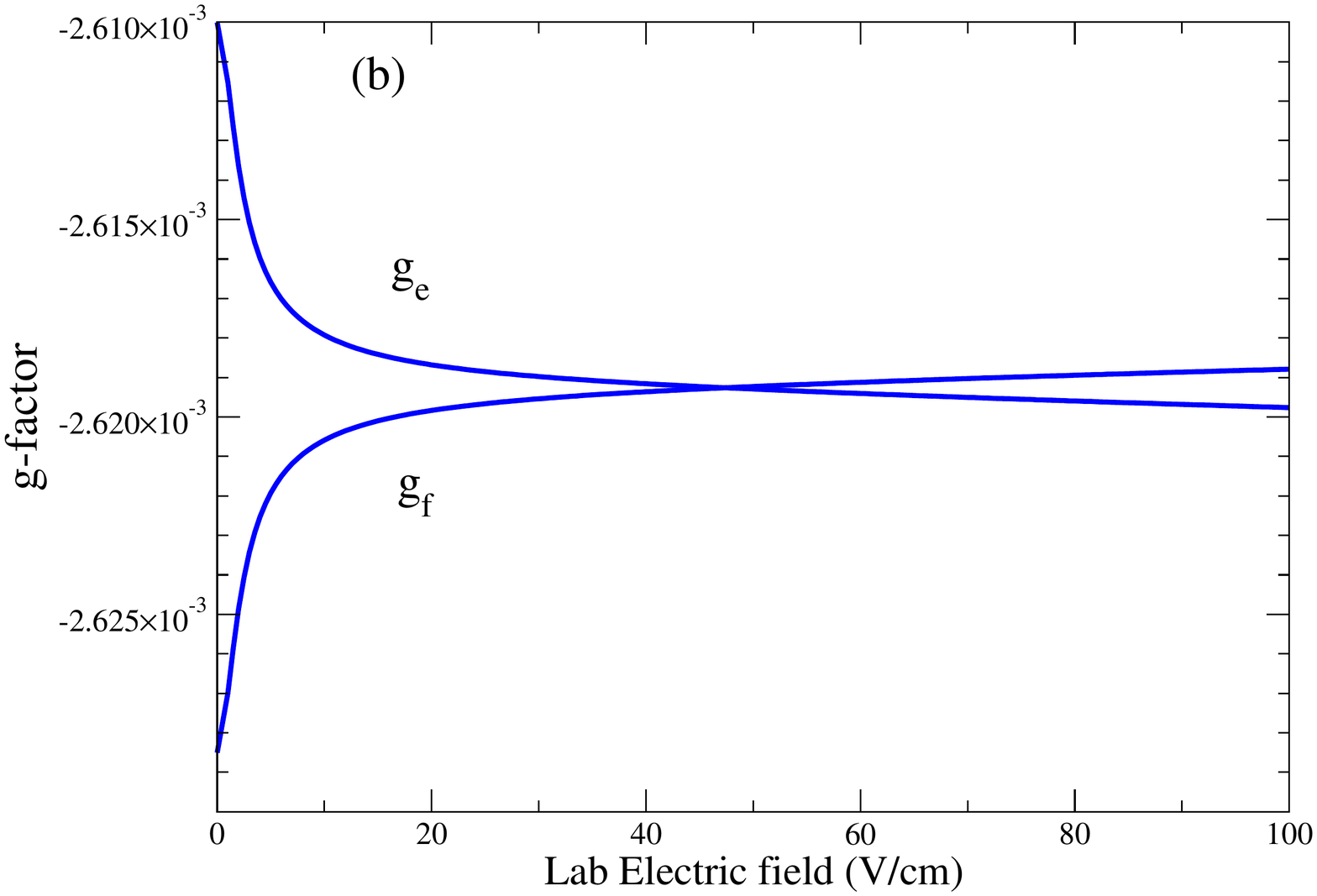}
 \caption{(Color online) Calculated $g_e$ and $g_f$ for $ H^3\Delta_1$ $^{232}$Th$^{16}$O as functions of the electric field. 
Both Zeeman and Stark interactions with the $^3\Delta_2$ and $^3\Pi_{0^\pm}$ states are taken into account.
Panel (a) $J=1, M=1$, Panel (b) $J=2, M=1$}
 \label{gfgecross}
\end{figure}

\begin{figure}
\begin{center}
\includegraphics[width = 3.3 in]{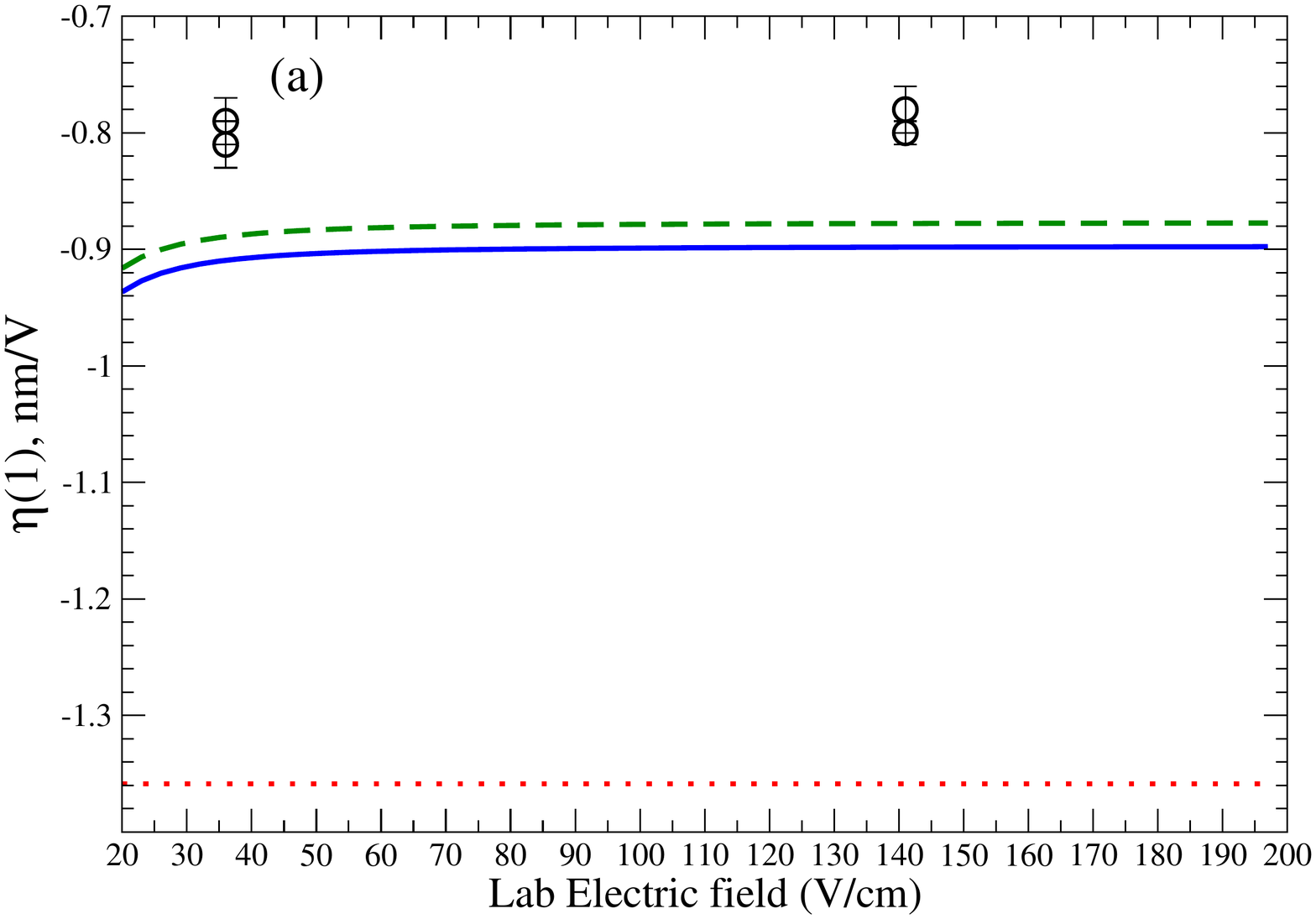}
\includegraphics[width = 3.3 in]{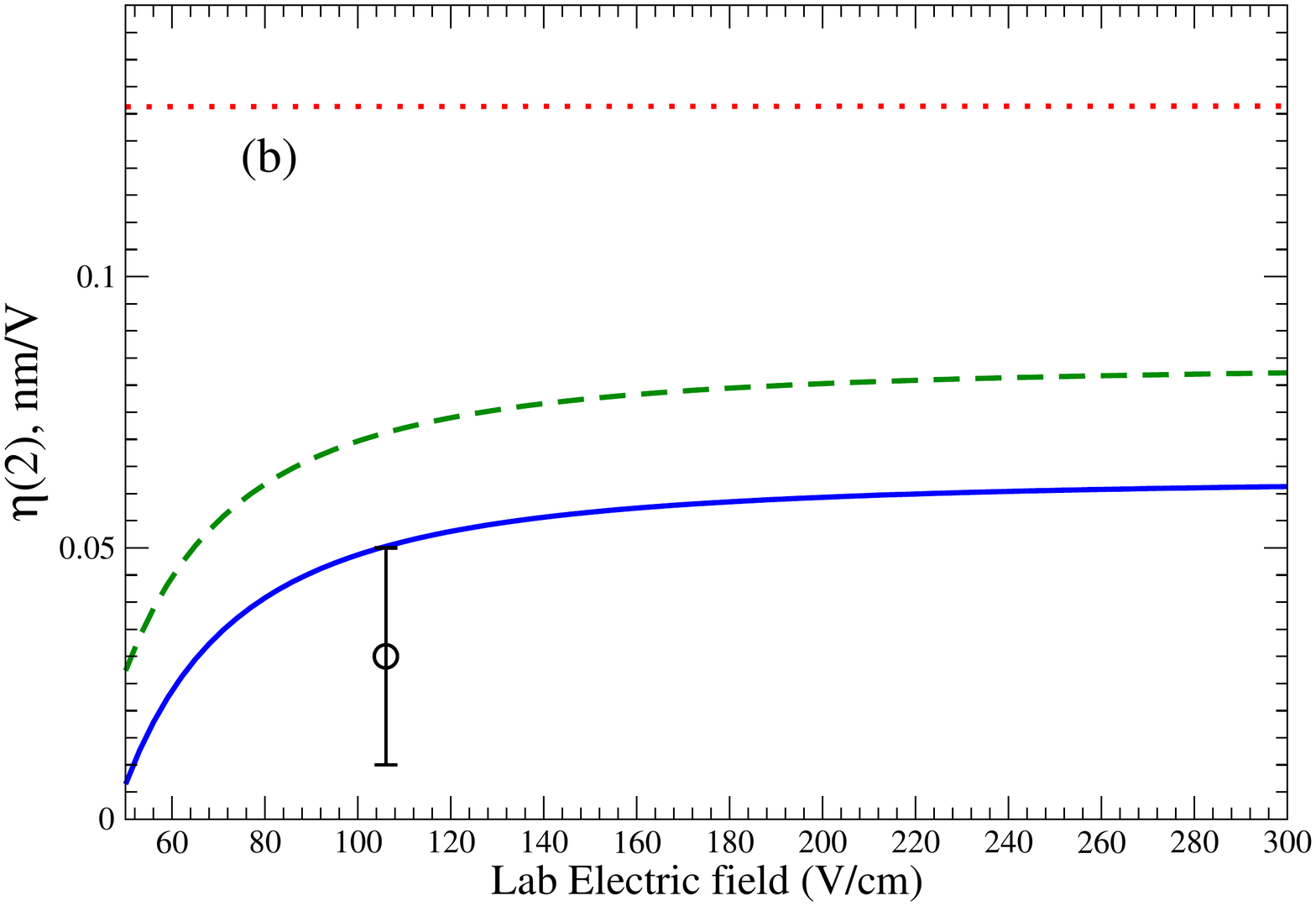}
\caption{\label{eta12} (Color online) Calculated $\eta(J)$ as functions of the electric field.
Solid (blue) curves - both Zeeman and Stark interactions with the $^3\Delta_2$ and $^3\Pi_{0^\pm}$ states are taken into account.
Dashed (green) curves - only the Zeeman interaction with $^3\Delta_2$ and $^3\Pi_{0^\pm}$ states is taken into account.
Dotted (red) curves - both Zeeman and Stark interactions with the $^3\Delta_2$ and $^3\Pi_{0^\pm}$ states are omitted.
Circles (black) - experimental values.
Panel (a) $J=1, M=1$, Panel (b) $J=2, M=1$}
\end{center}
\end{figure}

%%%%%%%%%%%%%%%%%%%%%%%%%%%%%%%%%%%%%%%%%%%%%%%%%%%%%%%%%%%%%%%%%%%%%%%%%%%%%%%
%%%%%%%%%%%%%%%%%%%%%%%%%%%%%%%%%%%%%%%%%%%%%%%%%%%%%%%%%%%%%%%%%%%%%%%%%%%%%%%

The PNPI$-$SPbU team acknowledge Saint-Petersburg State University for a research grant No.~0.38.652.2013 and RFBR Grant No.~13-02-01406. L.S.\ is also grateful to the President of RF grant No.~5877.2014.2 The molecular calculations were partly performed at the Supercomputer ``Lomonosov''.  The work of the Harvard and Yale teams was performed as part of the ACME Collaboration, to whom we are grateful for their contributions, and was supported by the NSF.

%\input{ThOg_22pt_bib_acme_edit.bbl}

%\bibliographystyle{./bib/apsrev}
%\bibliography{bib/JournAbbr,bib/SkripnikovLib,bib/QCPNPI,bib/TitovLib,bib/Kaldor,bib/PetrovLib,bib/Lomachuk,bib/ACME}

\end{document}